# Deconfined SU(2) vector fields at zero temperature

Bernd A. Berg

*Department of Physics, Florida State University, Tallahassee, FL 32306-4350, USA*

(Dated: June 22, 2010.)

Markov chain Monte Carlo simulations of pure SU(2)⊗U(1) lattice gauge theory show a (zero temperature) deconfining phase transition in the SU(2) gluon sector when a term is added to the SU(2) and U(1) Wilson actions, which requires joint U(2) gauge transformations of the SU(2) and U(1) vector fields. Investigations of this deconfined phase are of interest as it could provide an alternative to the Higgs mechanism.

The explicit breaking of gauge symmetry through massive vector bosons is avoided by the Higgs mechanism [1], which completes the standard electroweak theory [2] and ensures desired features such as renormalizability [3]. Nevertheless, the introduction of the Higgs particle remains quite adhoc and the discussion of alternatives is of interest as long as a Higgs boson has not been observed.

In Euclidean field theory notation the action of the electroweak gauge part of the standard model reads

$$S = \int d^4x \ \mathcal{L}, \quad \mathcal{L} = \frac{1}{4} F'_{\mu\nu} F'_{\mu\nu} + \frac{1}{2} \mathrm{Tr}\, F_{\mu\nu} F_{\mu\nu}, \quad (1)$$

$$F'_{\mu\nu} = \partial_\mu A'_\nu - \partial_\nu A'_\mu, \quad (2)$$

$$F_{\mu\nu} = \partial_\mu A_\nu - \partial_\nu A_\mu + ig \left[ A_\mu, A_\nu \right], \quad (3)$$

where $A'_\mu$ are U(1) and $A_\mu$ are SU(2) vector fields.

Typical textbook introductions of the standard model, e.g. [4], emphasize at this point that the theory contains four massless gauge bosons and introduce the Higgs mechanism as a vehicle to modify the theory so that only one gauge boson, the photon, stays massless. Such a presentation reflects that the introduction of the Higgs particle to electroweak interactions [1] preceded our nonperturbative understanding of non-Abelian gauge theories. In fact, there is already only the photon massless in the spectrum of [1]. The self interaction [3] of the SU(2) gauge part has dynamically generated a non-perturbative mass gap, and the SU(2) spectrum consists of massive glueballs. One of them can be used as mass scale at our disposal.

However, SU(2) gauge theory confines fermions, while the leptons are found as free particles. Coupling a Higgs field in the usual way causes a deconfining phase transition, so that fermions are liberated, a photon stays massless and glueballs break up into elementary massive vector bosons. Such a transition between the confinement and the Higgs phase has indeed been observed in pioneering lattice gauge theory (LGT) investigations [5].

The present study is in part motivated by the behavior of the Polyakov loop in put U(1) LGT. It is well known that U(1) LGT confines fermions in the strong coupling limit of its lattice regularization [6]. Towards weaker coupling it undergoes a transition into the Coulomb phase in which the effective potential for the Polyakov loop assumes the Mexican hat shape that is also characteristic for a scalar Higgs doublet in the broken phase. From simulations on a $12^4$ lattice with periodic boundary conditions Fig. 1 and 2 show for the U(1) Wilson action scatter plots of the Polyakov loop lattice averages from Monte Carlo (MC) simulations in the symmetric and in the broken phase, where the usual U(1) coupling is parametrized by $\beta' = 1/g_0'^2$. As this transition happens for U(1) LGT at zero temperature, the pictures hold for Polyakov loops winding in any one of the four directions through the torus (ordered starts are used to avoid metastabilities). The similarity with the behavior of a Higgs field is evident. At $\beta' = 2$ we are deep in the broken phase, while in our simulations below we shall use $\beta' = 1.1$ where one has less problems with U(1) metastabilities of the MC algorithm.

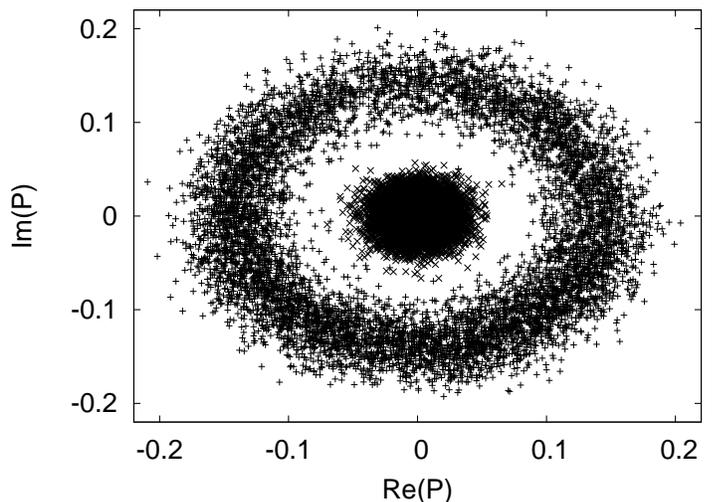

FIG. 1: Scatter plot for U(1) Polyakov loops on a $12^4$ lattice at $\beta' = 0.9$ in the symmetric phase (center) and at $\beta' = 1.1$ in the broken phase (ring).

One may like then to couple SU(2) gauge fields to the U(1) Polyakov loop, but a local operator is more attrac-

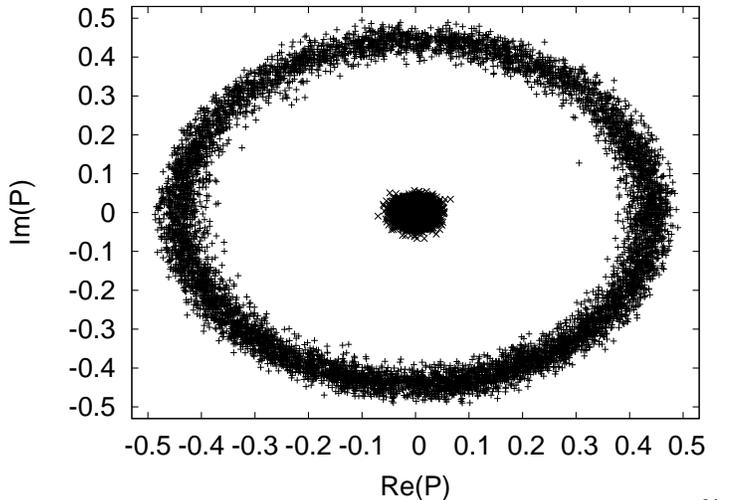

FIG. 2: Scatter plot for U(1) Polyakov loops on a $12^4$ lattice $\beta' = 0.9$ in the symmetric phase (center) and at $\beta' = 2$ in the broken phase (ring).

tive, so we substitute the U(1) gauge matrix itself. The simplest lattice action of the desired type is

$$
\begin{aligned}
S &= \beta'S' + \beta S + \lambda S^{\mathrm{add}} \\
&= \sum_p \left( \beta'S'_p + \beta S_p + \lambda S^{\mathrm{add}}_p \right) .
\end{aligned}
\tag{4}
$$

Here $S'$ and $S$ are the usual U(1) and SU(2) Wilson actions with bare coupling constants $\beta' = 1/g_0'^2$ and $\beta = 4/g_0^2$ [7]. The additional term is summed up from plaquette contributions of the form

$$
\frac{1}{2}\,\mathrm{ReTr}\left[ U'_\mu(x)\, U_\nu(x+\hat\mu)\, U'^*_\mu(x+\hat\nu)\, U^*_\nu(x) \right]
\tag{5}
$$

where U(1) matrices are denoted by $U'_\mu(x)$ and SU(2) matrices by $U_\mu(x)$. The $U'$ matrices are complex numbers, which may be multiplied with the $2 \times 2$ unit matrix when convenient. They commute with themselves and with the $U$ matrices. Notably, $S^{\mathrm{add}}$ requires common gauge transformations for the $U'$ and $U$ fields:

$$
U'_\mu(x) \to G(x)U'_\mu(x)G^{-1}(x+\hat\mu), \; U_\mu(x) \to G(x)U_\mu(x)G^{-1}(x+\hat\mu)
\tag{6}
$$

with $G = g'g$, $g' \in U(1)$, $g \in SU(2)$. This calls for an interpretation of the U(1) and SU(2) as U(2) subgroups. Let us recall at this point that the U(1) and SU(2) gauge theories are represented by the $S$ and $S'$ pieces of the action [4] instead of the SU(2)$\otimes$U(1) action, because the latter cannot be used, which has been suggested to indicate that the U(1) and SU(2) factor groups are broken down remnants of a larger simple group [8]. It is tempting to pursue a related interpretation of the gauge transformation encountered here.

In the gauge for which the SU(2) factor of the U(1) matrices is transformed to the unit matrix and the phase factor of the SU(2) matrices to 1, the classical continuum limit [5] becomes

$$
\lambda\,\mathrm{Tr}\,F^{\mathrm{add}}_{\mu\nu}F^{\mathrm{add}}_{\mu\nu}\,, \quad F^{\mathrm{add}}_{\mu\nu} = g_0\partial_\mu A_\nu - g'_0\partial_\nu A'_\mu\,.
\tag{7}
$$

This allows to couple matter fields with local U(2) invariance. Details are given in Ref. [9].

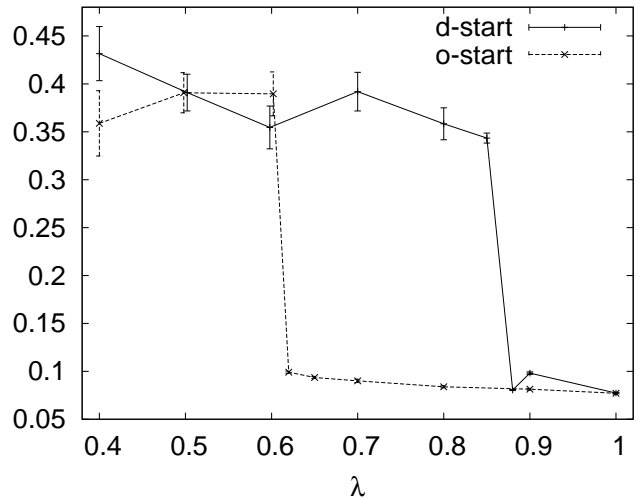

FIG. 3: SU(2) string tension from Creutz ratios on a $12^4$ lattice as function of $\lambda$ with $\beta' = 1.1$ and $\beta = 2.3$.

In the following results from MC simulations with the action [4] on $6^4$ and $12^4$ lattice with periodic boundary conditions are reported. Our MC procedure proposes the usual U(1) and SU(2) changes and relies on a variant of the biased Metropolis-heatbath algorithm [10]. In contrast to ordinary MC simulations of lattice gauge theories, updates stay within the fixed gauge introduced before Eq. [7]. The gauge freedom remaining within the MC procedure is a global transformation $U \to GUG^{-1}$ with the same $G$ for all $U$.

The U(1) coupling is kept at $\beta' = 1.1$ and for the SU(2) coupling the values $\beta = 2.2$, 2.3 and 2.4 are used. At $\lambda = 0$, without interaction, $\beta'$ is in the U(1) Coulomb phase and $\beta$ in the SU(2) scaling region. On the $12^4$ lattice Creutz ratios [11] for the SU(2) string tension $\kappa$ were calculated from Wilson loops up to size $5 \times 5$. Increasing $\lambda$, Fig. 3 shows their behavior at $\beta = 2.3$. Units with lattice spacing $a = 1$ are used. Data points are based on a statistics of at least $2^{10}$ sweeps without measurements and subsequently $32 \times 2^{10}$ sweeps with measurements. Error bars are calculated with respect to the 32 jackknife bins. Corresponding plots for $\beta = 2.2$ and $\beta = 2.4$ look very similar, only that a considerably larger statistics is needed for low $\lambda$ values at $\beta = 2.2$ to overcome the statistical noise.

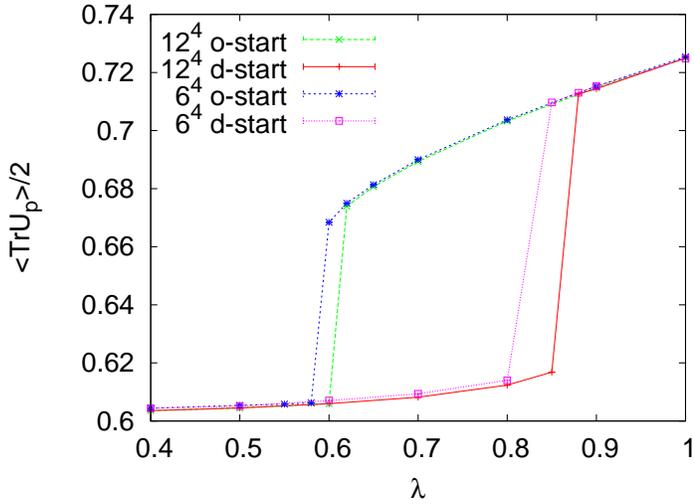

FIG. 4: SU(2) plaquette action on $6^4$ and $12^4$ lattices as function of $\lambda$ with $\beta' = 1.1$ and $\beta = 2.3$.

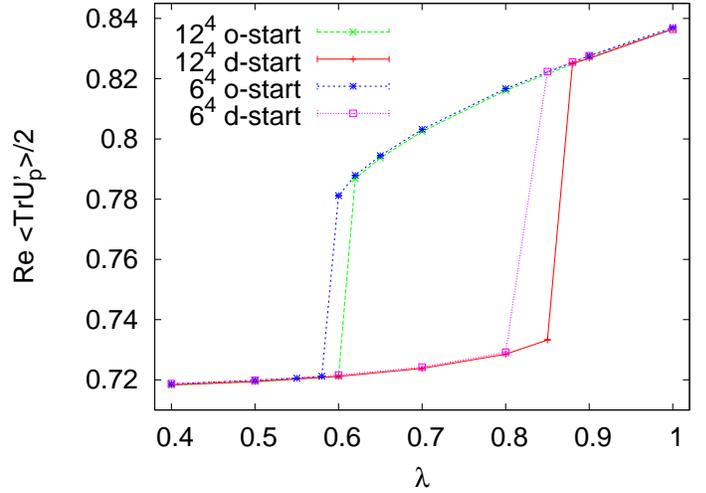

FIG. 6: U(1) plaquette action on $6^4$ and $12^4$ lattice as function of $\lambda$ with $\beta' = 1.1$ and $\beta = 2.3$.

slightly more ordered than the $12^4$, which is presumably due to the periodic boundary conditions. While the jump in the plaquette actions is similar for SU(2) and U(1), this is not the case for the string tensions. The SU(2) string tension decreases at the transition by a factor 3.5, whereas the drop for U(1) is just 25% of the original value (the metastability which is for both string tensions encountered for the disordered start at $\lambda = 0.9$ is not supposed to be of relevance).

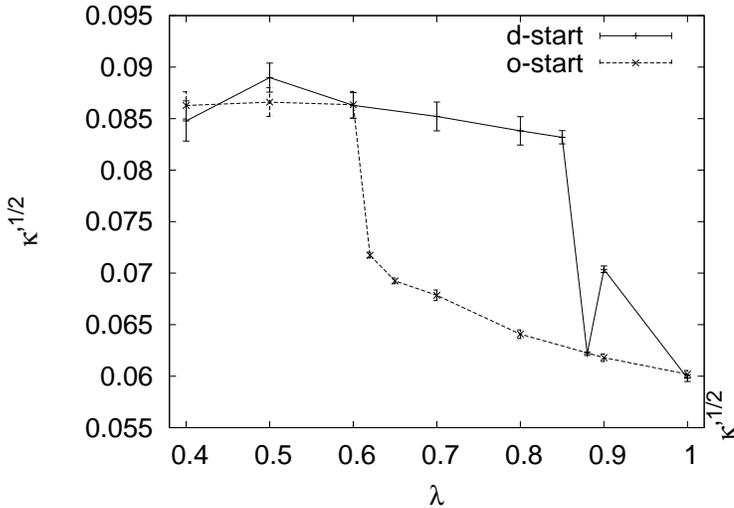

FIG. 5: U(1) string tension from Creutz ratios on a $12^4$ lattice as function of $\lambda$ with $\beta' = 1.1$ and $\beta = 2.3$.

Figure 3 shows a strong first order transition, which is also directly visible in the average SU(2) plaquette action $\langle \operatorname{Tr} U_p \rangle$ depicted in Fig. 4 and in the U(1) variables. See Fig. 5 for the U(1) string tension $\kappa'$ and Fig. 6 for the U(1) plaquette action. The $6^4$ lattice appears to be

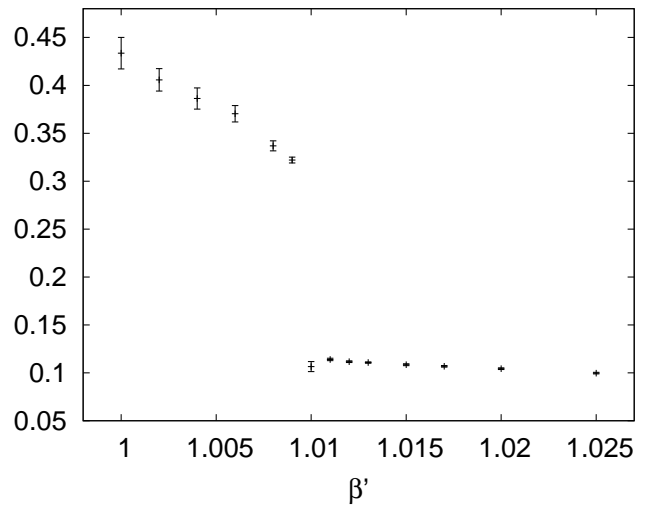

FIG. 7: U(1) string tension from Creutz ratios on a $12^4$ lattice as function of $\beta'$ ($\beta = \lambda = 0$).

The interpretation of our figures is that the U(1) string tension signals the deconfined phase on both sides of the transition, while the SU(2) string tension is characteristic for the confined phase at small $\lambda$ and for a zero-temperature deconfined phase at large $\lambda$. The latter point is supported by comparison with the behavior of the U(1) string tension for $\beta = \lambda = 0$ at the U(1) deconfining phase transition as shown in Fig. 7. The discontinuity in the string tension is as in Fig. 3 for SU(2), only that no strong metastabilities are observed for U(1) for which the opinion has remained that the transition is weakly first order as first reported in [12].

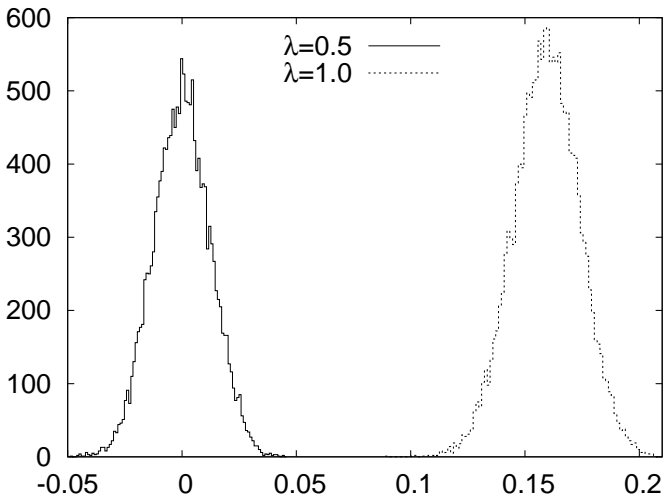

FIG. 8: SU(2) Polyakov loop histogram at $\beta' = 1.1$ and $\beta = 2.3$ with $\lambda = 0.5$ (left) and $\lambda = 1$ (right).

Polyakov loop measurements support also the SU(2) deconfining phase transition. For $\beta'$ and $\beta$ values as in the previous figures, Fig. 8 shows SU(2) Polyakov loop histograms at $\lambda = 0.5$ in the confined and at $\lambda = 1$ in the deconfined phase. In the confined phase the values scatter symmetrically about zero, whereas in the deconfined phase the $Z_2$ center symmetry is broken and the values scatter about a mean of 0.15930 (22). A very long run would produce a double peak, but the run time here was far too short to overcome the free energy barrier between the two peaks. Scatter plots of the U(1) Polyakov loops at the same couplings give approximately the ring of Fig. 1 at $\lambda = 0.5$ and the ring of Fig. 2 at $\lambda = 1$, i.e., both times deconfined.

With vanishing SU(2) string tension at sufficiently large $\lambda$ the SU(2) glueball spectrum is expected to break up into elementary vector bosons, which are conjectured to stay massive. Using irreducible representations of the cubic group on small Wilson loops, mass spectrum calculations were performed on lattices in the range $4^3 16$ to $12^3 64$ at selected coupling constant values. As in [13], correlation between U(1) plaquettes at momentum $k_1 = 2\pi/4$ of the $T_1^{+-}$ representation give clear evidence for a massless U(1) photon on both sides of the transition. Correlations between SU(2) loops indicate massive states on both sides of the transition with very weak signals in the deconfined phase. One can probe for vector boson masses by coupling fields in the fundamental SU(2) representations. As well-known conceptional problems [14] prevent one from putting left- or right-handed spinors on the lattice, and also Dirac fermions are very CPU time demanding, this may best be done by introducing a scalar doublet field. Gauge invariant trial operators for the $W$ boson mass are then given by [7]

$$W_{i,\mu}(x) = -i\,\mathrm{Tr}\,\left[\tau_i\,\alpha^\dagger(x+\hat\mu)\,U_\mu(x)\,\alpha(x)\right] \quad (8)$$

where $\tau_i$, $i = 1, 2, 3$ are the Pauli matrices and $\alpha(x)$ collects the angular variables of the scalar field. In contrast to Higgs model simulations on the lattice, these scalar fields should have positive squared masses and no self-interactions. They serve only as trial operators which allow one to test in a gauge invariant way for massive vector bosons. Assuming that the SU(2) gauge sector can be driven from a confined in a deconfined phase by the mixing with U(1) introduced in this paper, one would expect a degenerate $W$ triplet in the deconfined and scalar bound states in the confined phase, similarly as in Higgs model simulations. Such investigations will be the subject of future work.

To summarize, relying on LGT simulations, a SU(2) deconfining phase transition is obtained from interactions between the SU(2) and the U(1) vector fields. As this is achieved through spontaneous breaking of the SU(2) center symmetry one may expect that a renormalizable and asymptotically free continuum theory can be defined for suitable limits of the couplings $g_0'$, $g_0$ and $\lambda$. Though our model introduces a number of unusual features, the associated difficulties do not appear to be unsurmountable. Whether the observed transition can provide an alternative to the Higgs mechanism by generating a $W$ boson mass dynamically remains to be seen.

This work was in part supported by the DOE grant DE-FG02-97ER41022 and by a Humboldt Research Award at Leipzig University. I am indebted to Wolfhard Janke and his group for their kind hospitality. Further, I thank Arwed Schiller for useful discussions and Hao Wu for programing help. Some of the computer programs used rely on collaborations with Alexei Bazavov.